# Controlled Collisions for Multiparticle Entanglement of Optically Trapped Atoms

Olaf Mandel, Markus Greiner, Artur Widera, Tim Rom, Theodor W. Hänsch & Immanuel Bloch[*]

Sektion Physik, Ludwig-Maximilians-Universität, Schellingstrasse 4/III, D-80799 Munich, Germany & Max-Planck-Institut für Quantenoptik, D-85748 Garching, Germany.

**Entanglement lies at the heart of quantum mechanics and in recent years has been identified as an essential resource for quantum information processing and computation[1-4]. Creating highly entangled multi-particle states is therefore one of the most challenging goals of modern experimental quantum mechanics, touching fundamental questions as well as practical applications. Here we report on the experimental realization of controlled collisions between individual neighbouring neutral atoms trapped in the periodic potential of an optical lattice. These controlled interactions act as an array of quantum gates between neighbouring atoms in the lattice and their massively parallel operation allows the creation of highly entangled states in a single operational step, independent of the size of the system[5,6]. In the experiment, we observe a coherent entangling-disentangling evolution in the many-body system depending on the phase shift acquired during the collision between neighbouring atoms. This dynamics is indicative of highly entangled many-body states that present novel opportunities for theory and experiment.**

During the last years, Bose-Einstein condensates have been loaded into the periodic dipole force potential of a standing wave laser field – a so-called optical lattice. In these new systems, it has been possible to probe fundamental many body quantum mechanics in an unprecedented way, with experiments ranging from Josephson junction tunnel arrays[7,8] to the observation of a Mott insulating state of quantum gases[9,10]. Important applications for atoms in a Mott insulating state to quantum information processing have been envisaged early on. The Mott state itself, with one atom per lattice site, could act as a huge quantum memory, in which information would be stored in atoms at different lattice sites. Going far beyond these ideas, it has been suggested that controlled interactions between atoms on neighbouring lattice sites could be used to realise a massively parallel array of neutral atom quantum gates[5,11-14], with which a large multi particle system could be highly entangled[6] in a single operational step. Furthermore, the repeated application of the quantum gate array could form the basis for a universal quantum simulator along the original ideas of Feynman for a quantum computer as a simulator of quantum dynamics[15-17].

The basic requirement for such a unique control over the quantum state of a many body system including its entanglement is the precise microscopic control of the interactions between atoms on different lattice sites. In order to illustrate this, let us consider the case of two neighbouring atoms, initially in state $|\Psi\rangle = |0\rangle_j |0\rangle_{j+1}$ placed on the $j^{th}$ and $j+1^{th}$ lattice site of the periodic potential in the spin-state $|0\rangle$. First both atoms are brought into a superposition of two internal states $|0\rangle$ and $|1\rangle$ using a π/2 pulse such that $|\Psi\rangle = (|0\rangle_j + |1\rangle_j)(|0\rangle_{j+1} + |1\rangle_{j+1})/2$. Then a spin-dependent transport[18] splits the spatial wave packet of each atom such that the wave packet of the atom in state $|0\rangle$ moves to the left, whereas the wave packet of the atom in state $|1\rangle$ moves to the right. The two wave packets are separated by a distance $\Delta x=\lambda/2$, such that now $|\Psi\rangle = \left(|0\rangle_j|0\rangle_{j+1} + |0\rangle_j|1\rangle_{j+2} + |1\rangle_{j+1}|0\rangle_{j+1} + |1\rangle_{j+1}|1\rangle_{j+2}\right)/2$, where in the notation atoms in state $|0\rangle$ have retained their original lattice site index and $\lambda$ is the wavelength of the laser forming the optical periodic potential. The collisional interaction between the atoms[5,12,19] over a time $t_{hold}$ will lead to a distinct phase shift $\varphi = U_{01}t_{hold}/\hbar$, when both atoms occupy the same lattice site $j+1$ resulting in:

$|\Psi\rangle = \left(|0\rangle_j|0\rangle_{j+1} + |0\rangle_j|1\rangle_{j+2} + e^{-i\varphi}|1\rangle_{j+1}|0\rangle_{j+1} + |1\rangle_{j+1}|1\rangle_{j+2}\right)/2$. Here $U_{01}$ is the onsite-interaction matrix element that characterises the interaction energy when an atom in state $|0\rangle$ and an atom in state $|1\rangle$ are placed at the same lattice site and $\hbar$ is Planck's constant divided by $2\pi$. Alternatively a dipole-dipole interaction has been proposed[11] for generating a state dependent phase shift $\varphi$. The final many body state after bringing the atoms back to their original site and applying a last $\pi/2$ pulse can be expressed as

$$|\Psi\rangle = \frac{1+e^{-i\varphi}}{2}|1\rangle_j|1\rangle_{j+1} + \frac{1-e^{-i\varphi}}{2}|\text{BELL}\rangle.$$

Here $|\text{BELL}\rangle$ denotes the Bell-like state corresponding to

$\left(|0\rangle_j\left(|0\rangle_{j+1} - |1\rangle_{j+1}\right) + |1\rangle_j\left(|0\rangle_{j+1} + |1\rangle_{j+1}\right)\right)/2$.

This scheme can be generalised when more than two particles are placed next to each other, starting from a Mott insulating state of matter[9,10]. In such a Mott insulating state, atoms are localized to lattice sites, with a fixed number of atoms per site. For three particles e.g. one can show that if φ=(2n+1) π (with *n* being an integer), so called maximally entangled Greenberger-Horne-Zeilinger (GHZ) states[20] are realised. For a string of *N*>3 atoms, where each atom interacts with its left and right-hand neighbour (see Fig. 1), the entire string of atoms can be entangled to form so called cluster states in a single operational step[5,6]. The controlled interactions described above can be viewed as being equivalent to an ensemble of quantum gates acting in parallel[3,5].

The experimental setup used to load Bose-Einstein condensates into the three-dimensional optical lattice potential (see methods section) is similar to our previous work[10,19]. Briefly, we start with a quasi-pure Bose-Einstein condensate of $10^5$ $^{87}$Rb atoms in the $|F=1, m_F=-1\rangle$ state in a harmonic magnetic trapping potential with isotropic trapping frequencies of ω=2π×14 Hz. Here *F* and $m_F$ denote the total angular momentum and the magnetic quantum number of the atom's hyperfine state. The three-dimensional periodic potential of an optical lattice is then ramped up over a period of 80 ms to a potential depth of 25 $E_r$, such that the Bose-Einstein condensate is converted into a Mott insulating state. Here $E_r$ denotes the recoil energy $E_r=\hbar^2 k^2/2m$, with $k=2\pi/\lambda$ being the wave vector of the laser light and *m* the mass of a single atom. For our experimental parameters of atom number and harmonic confinement, such a Mott insulator should consist mainly of a central core with *n*=1 atom per lattice site[9,21,22]. The magnetic trapping potential is then rapidly switched off, but an actively stabilized magnetic offset field of 1 G along the transport direction is maintained to preserve the spin polarization of the atoms. With the optical standing wave along this direction we are able to realize a spin dependent transport of the atoms. After turning off the magnetic trapping field we wait another 40 ms for the electronics to stabilize the magnetic offset field. Thereafter, 3.5 ms before the quantum gate sequence is initiated, we adiabatically increase the lattice depth along this axis to 34 $E_r$ such that atoms remain in the vibrational ground state, are tightly confined and can be moved as fast as possible without excitations to higher vibrational states.

In the experiment, the two hyperfine states $|F=1, m_F=-1\rangle \equiv |0\rangle$ and $|F=2, m_F=-2\rangle \equiv |1\rangle$ form the logical basis of a single atom qubit at each lattice site. These two states can be coupled coherently using resonant microwave radiation around 6.8 GHz. A π/2 pulse allows us to place the atom in a coherent superposition of the two states within a time of 6 µs. After creating such a coherent superposition, we use a spin-dependent transfer to split and move the spatial wave function of the atom over half a lattice spacing in two opposite directions depending on its internal state (see Fig. 1). Such a movement process is carried out within a time of 40 µs in order to avoid any vibrational excitations[18] (the probability for excitations into higher lying vibrational states was measured to be less than 3%). Atoms on neighbouring sites then interact for a variable amount of time $t_{hold}$. After half of the hold time, a microwave π pulse is furthermore applied. This spin-echo type π pulse is mainly used to cancel unwanted single particle phase shifts e.g. due to inhomogeneities in the trapping potentials. It does not however affect the non-trivial and crucial collisional phase shift due to the interactions between the atoms. After such a controlled collision, the atoms are moved back to their original site. Then a final π/2 microwave pulse with variable phase α is applied and the atom number in state $|1\rangle$ relative to the total atom number is recorded.

The Ramsey fringes obtained in this way are shown in Fig. 2 for exemplary hold times $t_{hold}$ and for a wider range of hold times their visibility is plotted in Fig. 3. For short hold times, where no significant collisional phase shift is acquired, a Ramsey fringe with a visibility of approx. 50% is recorded. For longer hold times we notice a strong reduction in the visibility of the Ramsey fringe, with an almost vanishing visibility of approx. 5% for a hold time of 210 µs (see Fig. 2(b)). This hold time corresponds to an acquired collisional phase shift of φ=π for which we expect a minimum visibility if the system is becoming entangled.

For such an entangled state the probability for finding atoms in state $|1\rangle$ becomes independent of the phase α corresponding to a vanishing Ramsey fringe. This can be seen e.g. for the two particle case: when the phase α of the last pulse is kept variable, the maximally entangled state for a collisional phase φ = (2n+1) π can be expressed as:

$$|\Psi(\varphi=\pi)\rangle = \frac{1}{\sqrt{2}}(|0\rangle|-,\alpha\rangle + |1\rangle|+,\alpha\rangle), \text{ where } |-,\alpha\rangle \equiv \frac{1}{\sqrt{2}}(c_c^-|0\rangle - c_s^-|1\rangle) \text{ and}$$

$$|+,\alpha\rangle \equiv \frac{1}{\sqrt{2}}(c_s^+|0\rangle - c_c^+|1\rangle) \text{ with } c_c^\pm \equiv e^{\pm i\alpha}\cos\alpha \text{ and } c_s^\pm \equiv -(\pm i\sin\alpha - 1). \text{ Here the}$$

probability for finding an atom in either spin state e.g. $P(|1\rangle)$ is independent of α and equal to ½: $P(|1\rangle) = \frac{1}{8}\{|c_s^+|^2 + |c_s^-|^2 + 2\cdot|c_c^+|^2\} = \frac{1}{2}$. This indicates that no single particle operation can place all atoms in either spin-state when a maximally entangled state has been created. The disappearance of the Ramsey fringe has been shown to occur not only for a two-particle system, but is a general feature for an arbitrary N-particle array of atoms that have been highly entangled with the above experimental sequence[3,23]. A

vanishing Ramsey fringe can therefore in principle not distinguish between two-particle or multi-particle entanglement.

For longer hold times, the visibility of the Ramsey fringe increases again reaching a maximum of 55% for a hold time of 450 μs. Here the system becomes disentangled again, as the collisional phase shift is close to φ=2π and the Ramsey fringe is restored with maximum visibility.

The coherent "entanglement oscillations" of the many body system[6] are recorded for longer hold times by using the multi-particle interferometer sequence of Fig. 1(b), where the atoms are not brought back to their original site but are rather kept delocalized[18]. This allows us to observe the Ramsey fringe of the previous sequence as a spatial interference pattern in a single run of the experiment in analogy to a double-slit interference experiment, when a state selective time-of-flight detection is used. Images of such an interference pattern can be seen in Fig. 4 for different hold times $t_{hold}$. The coherent evolution again indicates the entangling-disentangling dynamics that the system undergoes for different collisional phase shifts φ (see Fig. 5).

Although the observed coherent dynamics in the vanishing and re-emergence of the Ramsey fringe does not provide a rigorous proof of a highly entangled multi-particle state, it is very indicative of such a state. So far, we cannot employ single atom measurement techniques to detect correlations between individual atoms in the cluster that would provide a quantitative measurement on the size of the entangled many-body state. It is clear however that the minimum visibility observed in the Ramsey fringes is dependent on the quality of our initial Mott insulating state and the fidelity of the quantum gate operations. In an ideal experimental situation with perfect fidelity for the multi particle quantum gates and a defect free Mott insulating state, this visibility should vanish for a phase shift of $\varphi=(2n+1)\pi$. For a finite fidelity of the quantum gates, caused e.g. by a 5% fractional error in the pulse areas of the microwave pulses, the minimum visibility would already increase to ~2%. If defects are present in the initial quantum state of the Mott insulator, e.g. vacant lattice sites, then the entangled cluster state will not extend beyond this vacancy and the visibility of the Ramsey fringe will become non-zero due to isolated atoms in the lattice. We have noticed for example that the quality of the Mott insulating state is deteriorated due to its prolonged uncompensated exposure to the potential gradient of gravity after the magnetic trapping potential is turned off. In addition to an imperfect creation of the Mott state, such vacancies could be caused by the superfluid shell of atoms surrounding the Mott insulating core[9,21,22] or spontaneous emission due to the laser light, which leads to excitations of approx. 5% of the atoms for our total experimental sequence times.

In our one-dimensional lattice shift the system is very susceptible to vacant lattice sites, as a defect will immediately limit the size of the cluster. However, the scheme can be extended to two- or three dimensions by using two additional lattice shift operations along the remaining orthogonal lattice axes. As long as the filling factor of lattice sites would exceed the percolation threshold (31% for a 3D simple cubic lattice system[24]) a large entangled cluster should be formed, making massive entanglement of literally 100000 of atoms possible in only three operational steps. For some of the applications of such a highly entangled state it will however be crucial to locate the position of the defects in the lattice.

For the future it will be fascinating to explore novel schemes for quantum computing that are based only on single particle operations and measurements on such a cluster state[2]. Here the large amount of entanglement in a cluster state can be viewed as a resource for quantum computations. But already now, even without the possibility of manipulating single atoms in the periodic potential, a quantum computer based on the controlled collisions demonstrated here could be able to simulate a wide class of complex Hamiltonians of condensed matter physics that are translationally invariant[12,17].

It is a pleasure to thank H. Briegel and I. Cirac for stimulating discussions, and A. Altmeyer and T. Best for experimental assistance. This work was supported by the EU under the QUEST program, the AFOSR and the "Bayerische Forschungsstiftung".

Correspondence and requests for materials should be addressed to I.B. (e-mail: imb@mpq.mpg.de).

**Methods**

**Optical lattices**

A three dimensional array of microscopic potential wells is created by overlapping three orthogonal optical standing waves at the position of the Bose-Einstein condensate. In our case the atoms are trapped in the intensity maxima of the standing wave light field due to the resulting dipole force[25,26]. The laser beams for two of the periodic potentials are operated at a wavelength of λ=820 nm with beam waists of approx. 210 μm at the position of the Bose-Einstein condensate. This gaussian laser beam profile leads to an additional isotropic harmonic confinement of the atoms with trapping frequencies of 40 Hz for lattice potential depths of 25 $E_r$. In this configuration, we populate almost 100000 lattice sites with an average atom number per lattice site of up to 1 in the centre of the lattice. The lattice structure is of simple cubic type, with a lattice spacing of λ/2 and oscillation frequencies in each lattice potential well of approx. 30 kHz for a potential depth of 25 $E_r$.

**State dependent lattice potentials**

Along a third orthogonal direction a standing wave potential at a wavelength of $\lambda_x$=785 nm is used, formed by two counter propagating laser beams with linear polarization vectors[5,11,18]. The angle θ between these polarization vectors can be dynamically adjusted through an electro-optical modulator and additional polarization optics. Such a lin-∠-lin polarization configuration can be decomposed into a $\sigma^+$ and a $\sigma^-$ polarized standing wave laser field, giving rise to potentials

$V_+(x,\theta) = V_0 \cos^2(k_x x + \theta/2)$ and $V_-(x,\theta) = V_0 \cos^2(k_x x - \theta/2)$. Here $V_0$ is the potential depth of the lattice. By changing the polarization angle θ one can control the separation $\Delta x = \theta/\pi \cdot \lambda_x / 2$ between the two potentials. When increasing θ, both potentials shift in opposite directions and overlap again for $\theta = n \cdot \pi$. For our experimental conditions, the dipole potential experienced by atoms in the $|1\rangle$ state is given by $V_-(x,\theta)$ and for atoms in the $|0\rangle$ state, it is dominated by the $V_+(x,\theta)$ potential[18]. For these laser beams, a waist of 150 μm has been used, resulting in a maximum potential depth of 34 $E_r$ and corresponding maximum vibrational trapping frequencies of 39 kHz.

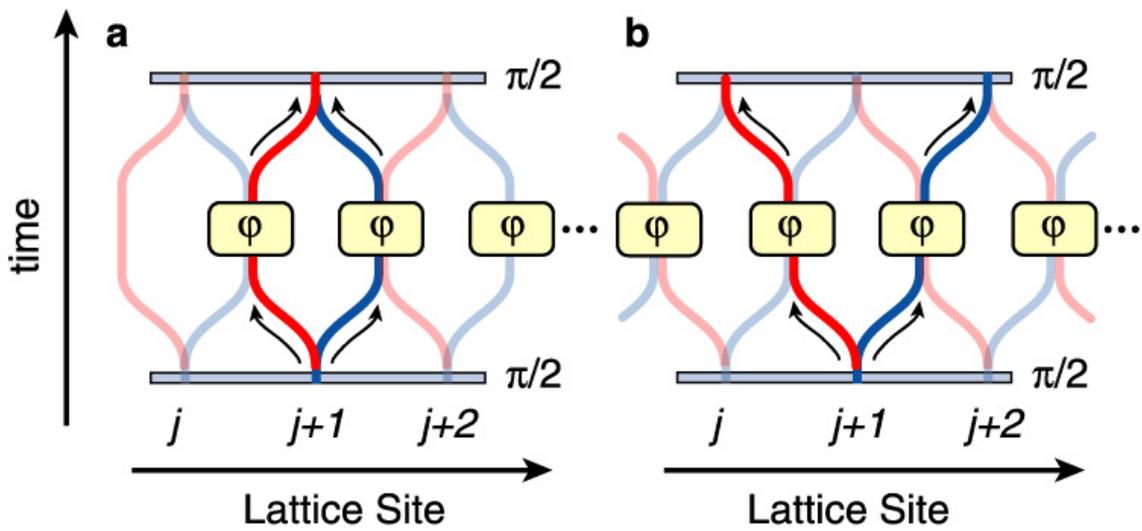

**Fig. 1.** Schematic multiple quantum gate sequences based on controlled interactions. **a** A chain of neutral atoms on different lattice sites is first placed in a coherent superposition of two spin states $|0\rangle$ (red) and $|1\rangle$ (blue) with a π/2 microwave pulse. Then a spin-dependent transport is used to split the spatial wave packet of an atom and move these two components along two opposite directions depending on their spin state. The wave packets are separated by a lattice period such that each atom is brought into contact with its neighbouring atom. Due to the collisional interaction between the atoms, a phase shift φ is acquired during a time $t_{hold}$ that the atoms are held on a common lattice site

depending on the spin state of the atoms. After such a controlled collisional interaction the wave packets of the individual atoms are returned to their original site and a final microwave $\pi/2$ pulse is applied to all atoms. This multiple quantum gate sequence can be equivalently described as a controllable quantum Ising interaction[6,12]. **b** In a slight modification of such a sequence, the atoms are not returned to their original lattice site $j+1$ but rather delocalized further over the $j^{th}$ and $j+2^{th}$ lattice site after the controlled collisional interaction. The small arrows indicate the different paths that a single atom will follow during the multiple quantum gate sequence. Both sequences can be viewed as multi-particle interferometers, where the many-body output state of the interferometer can in general not be expressed as a product state of single particle wavefunctions.

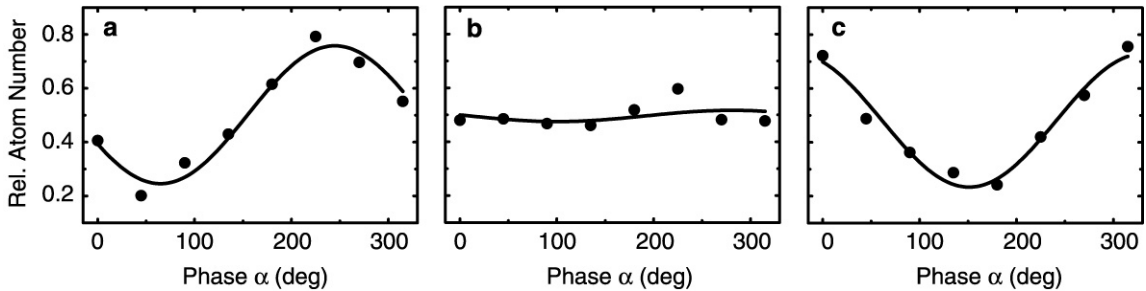

**Fig. 2** Experimentally measured Ramsey fringes for different hold times $t_{hold}$ during which atoms undergo a controlled collisional interaction with their neighbouring atoms. The experimental sequence used is similar to the one in Fig. 1a, where atoms are returned to their original lattice site after the controlled interaction. The hold times $t_{hold}$ are **a** 30 µs, **b** 210 µs and **c** 450 µs. The relative number of atoms $N_{rel}=N_1/N_{tot}$ in the $|1\rangle$ state vs the phase $\alpha$ of the final microwave $\pi/2$ pulse is measured. A state selective absorption imaging of the atom cloud is used to obtain $N_1$ after a time of flight period of 12 ms and 110 µs thereafter the total atom number is measured to yield $N_{tot}$. The solid line indicates a fit of a sinusoidal function with variable amplitude and an offset to the data from which the visibility of the Ramsey fringe is extracted. The change in the phase of the Ramsey fringes for different hold times is mainly caused by the different exposure times of the two spin-states of an atom to differential light shifts of the optical lattice that are not perfectly cancelled in the spin-echo sequence.

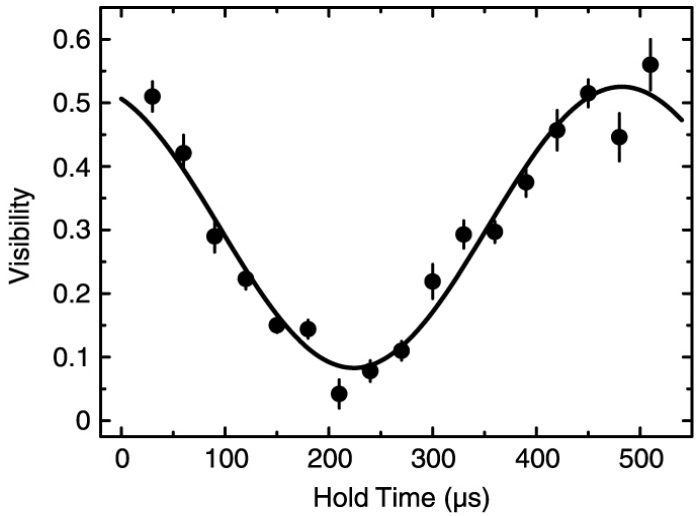

**Fig. 3** Visibility of Ramsey fringes vs. hold times on neighbouring lattice sites for the experimental sequence similar to the one displayed in Fig. 1a. The solid line is a sinusoidal fit to the data including an offset and a finite amplitude. Such a sinusoidal behaviour of the visibility vs. the collisional phase shift (determined by the hold time $t_{hold}$) is expected for a Mott insulating state with an occupancy of $n=1$ atom per lattice site[23]. The maximum observed visibility is limited to 55% by inhomogeneities and time dependent fluctuations of the lattice potentials

throughout the cloud of atoms that are not perfectly compensated by the applied spin-echo sequence (see text).

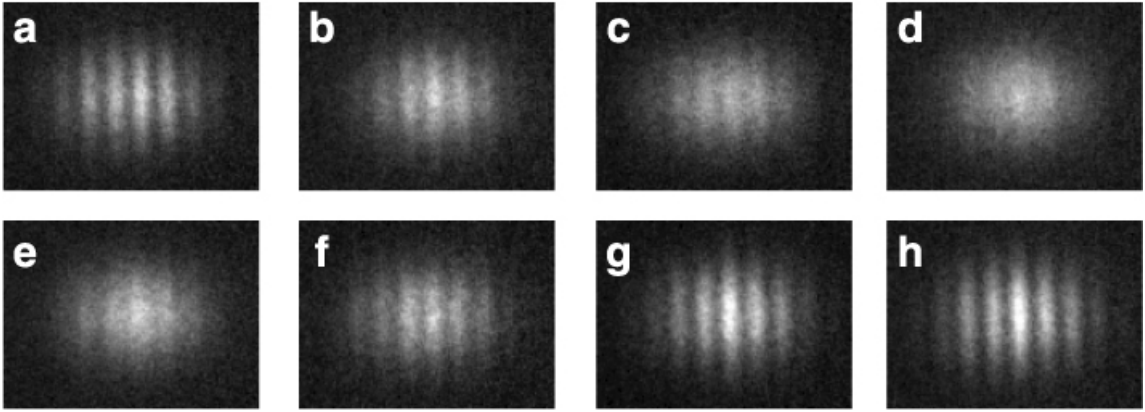

**Fig. 4** Spatial interference patterns recorded after applying the multiple quantum gate sequence of Fig. 1b for different collisional interaction times $t_{hold}$. The different hold times of **a** 30 μs, **b** 90 μs, **c** 150 μs, **d** 210 μs, **e** 270 μs, **f** 330 μs, **g** 390 μs and **h** 450 μs lead to different collisional phase shifts φ, ranging from approx. $\varphi \approx 0$ **(a)** to just over $\varphi \approx 2\pi$ **(h)**. The vanishing and reappearance of the interference pattern is caused by the coherent entangling-disentangling dynamics in the many body system due to the controlled collisions between neighbouring atoms. The state selective absorption images were obtained after a time of flight period of 11 ms.

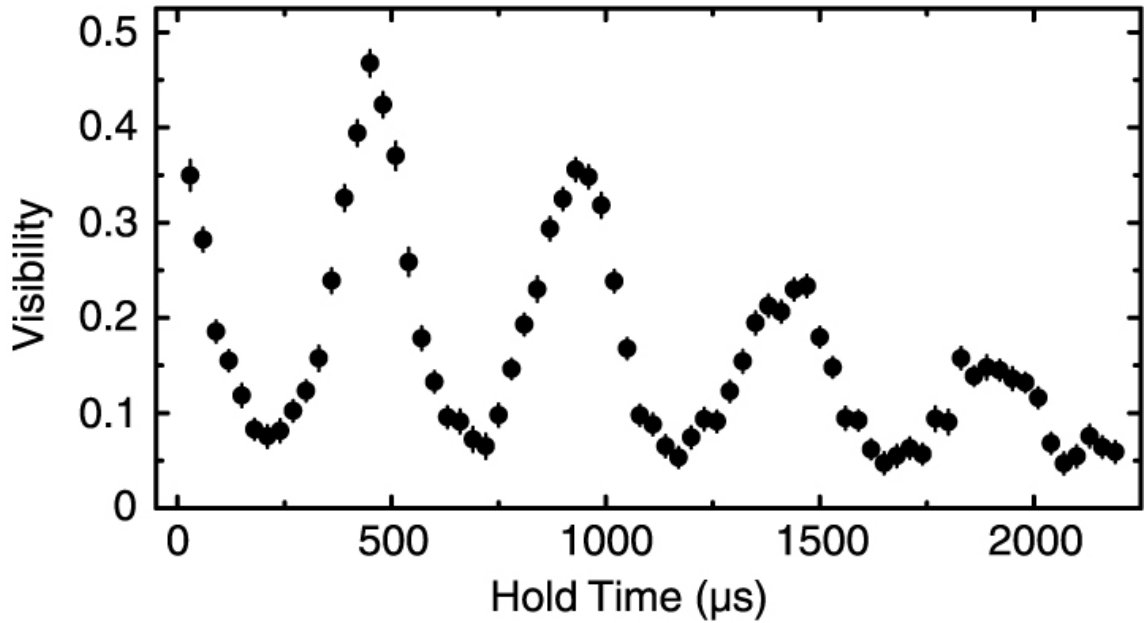

**Fig. 5** Visibility of the spatial interference patterns vs. different collisional interaction times $t_{hold}$. We have been able to observe up to four entangling-disentangling cycles in the experiment. The reduced visibility for longer hold times is mainly caused by a dephasing over the trapped cloud of atoms due to inhomogeneities in the external potentials.